\documentclass{jpc} %%% last changed 2014-08-20


\usepackage{natbib} 
\usepackage[ruled]{algorithm2e}
\theoremstyle{plain} %\crefname{satz}{Satz}{S\"atze}

\def\eg{e.g.}

\usepackage{booktabs}
\usepackage{epsfig,endnotes,xspace,url,diagbox}
\usepackage{amssymb, amsmath, amsthm, amsfonts}
\usepackage{enumerate}
\usepackage{hyperref}
\usepackage{varioref}
\usepackage{graphicx}
\usepackage[font=footnotesize]{subfig}
\usepackage{epstopdf}
\usepackage{xcolor}
\usepackage{multirow}
\usepackage{comment}
\usepackage[utf8]{inputenc}
\usepackage{mathtools}
\usepackage{wrapfig}
\usepackage{tabularx}
\usepackage{xspace}

\usepackage{cases}
\usepackage{stfloats}
\usepackage{fancyhdr}
\usepackage{enumitem}
\usepackage{caption}

\definecolor{revision}{RGB}{0,0,0}
\renewcommand{\revision}[1]{#1 \xspace}
\newcommand{\revisionstart}{\begin{color}{revision}}
\newcommand{\revisionend}{~\!\!\end{color}}

\newcommand{\red}[1]{{\textcolor{red}{#1}}}
\newcommand{\blue}[1]{{\textcolor{blue}{#1}}}
\newcommand{\brown}[1]{{\textcolor{brown}{#1}}}

\newcommand{\mypara}[1]{\smallskip\noindent\textbf{#1.} \xspace}

\newcommand{\etc}{{etc.}\xspace}

\renewcommand{\Pr}[1]{\ensuremath{\mathsf{Pr}\left[#1\right]}\xspace}
\newcommand{\myexp}[1]{\ensuremath{e^{#1}}\xspace}

\newcommand{\Data}{\ensuremath{D}\xspace}

\newtheorem{definition}{Definition}

\newtheorem{example}{Example}

\newcommand{\synflow}{\ensuremath{\mathsf{MCF}}\xspace}

\newcommand{\mcf}{\ensuremath{\mathsf{MCF}}\xspace}

\newcommand{\gum}{\ensuremath{\mathsf{GUM}}\xspace}

\newcommand{\margselect}{\ensuremath{\mbox{InDif}}\xspace}
\newcommand{\margselectfull}{Independent Difference\xspace}

\renewcommand{\AA}{\mathcal{A}\xspace}
\newcommand{\calN}{\mathcal{N}\xspace}

\newcommand{\ds}{\ensuremath{D_s}\xspace}

\newcommand{\zcdp}{zCDP\xspace}
\newcommand{\Lapp}[1]{\ensuremath{\mathcal{L}\left(#1\right)}\xspace}

\begin{document}

\title{DPSyn: Experiences in the NIST Differential Privacy Data Synthesis Challenges}

\sloppypar

\author{Ninghui Li}	%required
\address{Department of Computer Science, Purdue University}	%required
\email{ninghui@cs.purdue.edu}  %optional
% %\thanks{thanks 1, optional.}	%optional

\author{Zhikun Zhang}	%optional
\address{Control Science Department, Zhejiang University; CISPA Helmholtz Center for Information Security}	%optional
\email{zhikun.zhang@cispa.de}  %optional
\thanks{Work done while Zhikun Zhang was a visiting student at Purdue University.}	%optional

\author{Tianhao Wang}	%optional
\address{Department of Computer Science, Purdue University}
\email{tianhaowang@purdue.edu}

\begin{abstract}
    \noindent We summarize the experience of participating in two differential privacy competitions organized by the National Institute of Standards and Technology (NIST).  
    In this paper, we document our experiences in the competition, the approaches we have used, the lessons we have learned, and our call to the research community to further bridge the gap between theory and practice in DP research.
\end{abstract}

\maketitle

%% start the paper here:
\section{Introduction}
\label{sec:introduction}

In 2018-2019, the Public Safety Communications Research (PSCR) Division at the National Institute of Standards and Technology (NIST) ran two challenges regarding differential privacy.  
NIST's prominent history in using competitions to select the Advanced Encryption Standard (AES) algorithm and various Secure Hash Algorithm (SHA) algorithms immediately provide a high level of credibility and incentive for participants.  Our research group at Purdue University has been conducting research on data privacy for over a decade at that point.  We have often observed that theoretical utility bounds on DP algorithms are often not good indicators of their empirical performances.
While any research group can perform empirical comparisons with other approaches, such comparisons cannot be authoritative, since there are often a large number of parameters that one can choose in any empirical comparisons.  
We thus enthusiastically participated in the challenges.

In the first challenge, called the ``Unlinkable Data Challenge: Advancing Methods in Differential Privacy'',  contestants submit concept papers proposing a mechanism to enable the protection of personally identifiable information while maintaining a dataset's utility for analysis.  We developed an approach that extends our previous algorithm on publishing $k$-way marginals called PriView~\cite{qardaji2014priview}.  In our approach, one first obtains multiple marginal tables in a way that satisfies DP, then uses the method in~\cite{qardaji2014priview} to make them consistent, and finally synthesizes data based on these marginals.  Our concept paper won 2nd place, with the first place went to a proposal using private Generative Adversarial Network (GAN).  While GAN is a powerful and intriguing technique for learning generative models in domains such as images, our experiences tell us that using GAN is unlikely to outperform marginal-based approaches on relational data, as marginals are arguably the most privacy-efficient way to extract information from relational datasets.

In the second challenge, the ``Differential Privacy Synthetic Data Challenge'', participants implement their designs and empirically evaluate their artifacts on real datasets.  The 2nd challenge is organized in three rounds, each lasting about one month.  In each round, a sample dataset is given, and one submits dataset synthesized under different privacy parameters.  Top teams are then invited to submit the code for private synthesis, which are tested on another dataset that is similar in nature.  Our implementation won the 2nd place in all the three rounds.  The first two rounds were won by a team of two participants who appear to be software developers, and the last round was won by Ryan McKenna using approaches documented in~\cite{mckenna2019graphical}.  
NIST collected algorithm descriptions from top finishers of the final round.  The approaches from the top 4 teams all use marginals.  They differ in how to select marginals, which marginals to use, and how to use the marginals to synthesize data.

After the competition, we reflected upon the effort in the manual marginal engineering, and developed an algorithm to privately select marginals, and further refined and evaluated our data synthesis algorithm.  We documented these results in a paper that has been accepted to appear in 2021 USENIX Security symposium, and a version of it is available at~\cite{zhang2021privsyn}.  For discussions of the data synthesis algorithm, and related work to our approach, we refer the readers to~\cite{zhang2021privsyn}.

The rest of this paper is organized as follows. 
In Section~\ref{sec:back}, we review background on publishing marginals under DP.  
In Section~\ref{sec:prior}, we describe relevant research experiences and results prior to the competition that enabled us to participate in the competition.  
In Section~\ref{sec:dpsyn}, we summarize our submission to the first challenge.  
In Section~\ref{sec:setup}, we describe our approach to the second challenge. 
In Section~\ref{sec:gap}, we discuss the gap between theoretical analysis and empirical experiments in the design of differentially private mechanisms.
We conclude the paper in Section~\ref{sec:conclusion}.

\section{Background}
\label{sec:back}

\subsection{Marginal Table}
\begin{figure}[t]
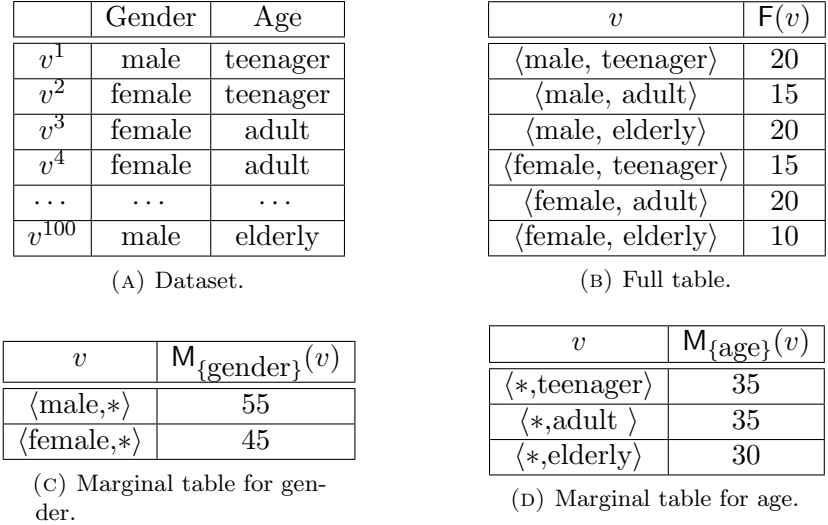

	\subfloat[Dataset.]{
		\begin{minipage}[t]{0.4\textwidth}
			\centering
			\begin{tabular}[c]{|c|c|c|}
				\hline \label{table:dataset}
				& Gender    & Age          \\ \hline\hline
				$v^1$       & male      & teenager     \\ \hline
				$v^2$       & female    & teenager     \\ \hline
				$v^3$       & female    & adult        \\ \hline
				$v^4$       & female    & adult        \\ \hline
				$\cdots$    & $\cdots$  & $\cdots$    \\ \hline 
				$v^{100}$       & male      & elderly      \\ \hline
			\end{tabular}
		\end{minipage}
	}
	\subfloat[Full table.]{
		\begin{minipage}[t]{0.4\textwidth}
		\centering
			\begin{tabular}[c]{|c|c|}
				\hline \label{table:f}
				$v$					& $\mathsf{F}(v)$ \\ \hline\hline
				$\langle$male, teenager$\rangle$    & 20    \\ \hline
				$\langle$male, adult$\rangle$       & 15      \\ \hline
				$\langle$male, elderly$\rangle$     & 20    \\ \hline
				$\langle$female, teenager$\rangle$  & 15    \\ \hline
				$\langle$female, adult$\rangle$     & 20  \\ \hline 
				$\langle$female, elderly$\rangle$   & 10      \\ \hline
			\end{tabular}
		\end{minipage}
	} \\
	\subfloat[Marginal table for gender.]{
		\begin{minipage}[t]{0.4\textwidth}
			\centering
			\begin{tabular}[c]{|c|c|}
				\hline \label{table:m1}
				$v$					& $\mathsf{M_{\{\mbox{gender}\}}}(v)$ \\ \hline\hline
				$\langle$male,$*\rangle$    & 55    \\ \hline
				$\langle$female,$*\rangle$  & 45      \\ \hline
			\end{tabular}
		\end{minipage}
	}
	\subfloat[Marginal table for age.]{
		\begin{minipage}[t]{0.4\textwidth}
			\centering
			\begin{tabular}[c]{|c|c|}
				\hline \label{table:m2}
				$v$					& $\mathsf{M_{\{\mbox{age}\}}}(v)$ \\ \hline\hline
				$\langle*$,teenager$\rangle$    & 35    \\ \hline
				$\langle*$,adult   $\rangle$    & 35      \\ \hline
				$\langle*$,elderly$\rangle$     & 30      \\ \hline
			\end{tabular}
		\end{minipage}
	}
    \caption{Example of the dataset, the full table, and two marginal tables.}
    \label{fig:example_marginal}
\end{figure}

Marginal tables capture the correlations among a set of attributes.  Given a dataset, a marginal table provides the synopsis of the dataset summarized on a subset of attributes.  Figure~\ref{fig:example_marginal} shows an example dataset and some marginals computed from it.  
Marginal tables can be computed with low degree of noises while satisfying differential privacy, which we discuss below.

\subsection{Differential Privacy}

Intuitively, the notion of Differential Privacy (DP)~\cite{DMNS06} requires that any single record in a dataset has only a limited impact on the output.

\begin{definition}[$(\epsilon,\delta)$-Differential Privacy] \label{def:non-pure-dp}
	A randomization algorithm $\AA$ satisfies $(\epsilon,\delta)$-differential privacy ($(\epsilon,\delta)$-DP), where $\epsilon>0, 0\le \delta < 1$,
	if and only if for any two neighboring datasets $D$ and $D'$ that differ in one record, we have
	\begin{equation*}
	\forall{T\subseteq\! \mathit{Range}(\AA)}:\; \Pr{\AA(D)\in T} \leq e^{\epsilon}\, \Pr{\AA(D')\in T}+\delta,\label{eq:npdp}
	\end{equation*}
	where $\mathit{Range}(\AA)$ denotes the set of all possible outputs of the algorithm $\AA$.
\end{definition}

One way to understand DP is that for each individual whose data is included in the dataset $D$, one considers a hypothetical world in which the individual's data is removed from $D$ (i.e., a world in which $D'$ instead of $D$ is used as the input).  Since this hypothetical world does not contain private data about the individual, it is considered as an \underline{idealized world of privacy} for the individual.  While information regarding this individual may still be leaked due to correlation, leakage is not due to usage of the individual's data.  Satisfying $(\epsilon,\delta)$-DP means simulating the idealized worlds for all individuals simultaneously.  More specifically, if an algorithm $\AA$ satisfies $(\epsilon,0)$-DP, then any output $\AA(D)$ could also occur in the idealized world $D'$, albeit with a different probability (with probability ratio bounded by $e^\epsilon$).  If an algorithm $\AA$ satisfies $(\epsilon,\delta)$-DP, then there may exist bad outcomes for which leakage of some individual's information is not bounded by $e^\epsilon$; however, the probability that such outcomes occur is at most $\delta$.  Note that to use this justification for DP, all the data from one individual should be contained in a single record. 

There are two different ways of defining when two datasets $D$ and $D'$ are \textbf{neighboring}.  One is to define it as  $D'$ can be obtained from $D$ by \emph{either adding or removing one record}, which results in what is called \underline{Unbounded Differential Privacy}.  The other is to define neighboring as $D'$ can be obtained from $D$ by \emph{changing the value of exactly one record}, which results in \underline{Bounded Differential Privacy}.  From the point of view that DP simulates idealized worlds, both definitions are acceptable.  In Unbounded DP, the idealized world for an individual is achieved by removing the individual's data.  In Bounded DP, this is achieved by overwriting one individual's data (e.g., with some default values).

\subsection{Basic Primitives}

We now give background on the basic primitives for satisfying DP, and their applications to publish marginals. 

\mypara{Laplace Mechanism}
The Laplace mechanism computes a function $f$ on input dataset $D$ while satisfying $(\epsilon,0)$-DP, by adding to $f(D)$ a random noise.  The magnitude of the noise depends on $\mathsf{GS}_f$, the \emph{global $L_1$ sensitivity} of $f$, defined as,
\[
\mathsf{GS}_f = \max\limits_{D\simeq D'} ||f(D) - f(D')||_1
\] 
When $f$ outputs a single element, the Laplace mechanism is given below:
$$
\begin{array}{crl}
& \AA_f(D) & =f(D) + \Lapp{\frac{\mathsf{GS}_f}{\epsilon}}
\end{array}
$$
In the definition above, $\Lapp{\beta}$ denotes a random variable sampled from the Laplace distribution with scale parameter $\beta$; that is, $\Pr{\Lapp{\beta}=x} = \frac{1}{2\beta} \myexp{-|x|/\beta}$.  
When $f$ outputs a vector, $\AA$ adds independent samples of $\Lapp{\frac{\mathsf{GS}_f}{\epsilon}}$ to each element of the vector.  The variance of each such sample is $\frac{2 \mathsf{GS}_f^2}{\epsilon^2}$. 

A marginal table is a function that outputs a vector of record counts.  Its global $L_1$ sensitivity is $1$ under Unbounded DP (adding or removing one record can change the value of at most one cell by one), and $2$ under Bounded DP (changing one record could decrease the count of one cell by $1$ and increase the count of another by $1$). 

\mypara{Gaussian Mechanism}
Instead of adding Laplace-distributed noises, one could also add Gaussian noises, for which the magnitude of the noise depends on $\Delta_f$, the \emph{global $L_2$ sensitivity} of $f$.  Such a mechanism $\AA$ is given below:
$$
\begin{array}{crl}
& \AA(D) & =f(D)+\calN\left(0, \Delta_f^2 \sigma^2 \mathbf{I} \right)
  \\
\mbox{where} &  \Delta_f & = \max\limits_{(D,D') : D \simeq D'} || f(D) - f(D')||_2.
\end{array}
$$

In the above, $\calN (0, \Delta_f^2 \sigma^2 \mathbf{I})$ denotes a multi-dimensional random variable sampled from the normal distribution with mean $0$ and standard deviation  $\Delta_f \sigma$.  The global $L_2$ sensitivity for publishing a marginal is $1$ under Unbounded DP, and $\sqrt{2}$ under Bounded DP.  

It is known~\cite{dpbook} that for \revision{any $\epsilon<1$ and $\delta \in (0, 1)$}, when $\sigma=\sqrt{2\ln\frac{1.25}{\delta}}/\epsilon$, the Gaussian mechanism satisfies $(\epsilon,\delta)$-DP.  The variance of each such noise is $\frac{2 \Delta_f^2 \ln(1.25/\delta)}{\epsilon^2}$.  

When one needs to publish just one marginal, it is better to use the Laplace mechanism, since the variance is smaller.  When using Gaussian mechanism to satisfy the same $\epsilon$ and a reasonably small $\delta$, the variance is increased by a factor of $\ln(1.25/\delta)$.

\subsection{Composition of DP Mechanism}
\label{subsec:composition}

When we need to publish multiple marginals while satisfying DP, we need to analyze the effect of composing multiple DP mechanisms.  Here, there are several tools that we can use. 

\mypara{Basic Sequential Composition}
Using the standard sequential composition result, one can simply add the $(\epsilon,\delta)$ values in composition.  More specifically, given $k$ mechanisms $\mathcal{A}_1, \ldots, \mathcal{A}_k$ satisfying $(\epsilon_i, \delta_i)$-DP for $i=1,\ldots, k$ respectively, publishing the outputs of all these mechanisms satisfies $(\sum_i^k\epsilon_i, \sum_i^k\delta_i)$-DP.  Using the guarantee of the basic composition and allocating the privacy budget $(\epsilon,\delta)$ equally, publishing $k$ marginals results in each one having $\frac{1}{k}$ of the total budget. 

\mypara{Advanced Composition}
The advanced composition bound from~\cite{DRV10} states that the composition of $k$ mechanisms that each satisfies $(\epsilon,\delta)$-DP, satisfies ($\epsilon \sqrt{2k\log(1/\delta')} + k \epsilon (e^\epsilon-1)$, $k\delta+\delta'$)-DP, for any $\delta\in (0,1)$.  

\mypara{Zero Concentrated DP}
The notion of zero Concentrated Differential Privacy (\zcdp for short) offers elegant composition properties with tighter bounds. The general idea is to connect $(\epsilon, \delta)$-DP to R\'enyi divergence, and then use the properties of R\'enyi divergence to achieve tighter composition property.
Formally, \zcdp is defined as follows:

\begin{definition}[Zero-Concentrated Differential Privacy (\zcdp)~\cite{bun2016concentrated}]
A randomized mechanism $\mathcal{A}$ is $\rho$-zero concentrated differentially private (i.e., $\rho$-\zcdp) if for any two neighboring databases $D$ and $D'$ and all $\alpha \in (1, \infty)$,
\begin{align*}
\mathcal{D}_\alpha (\mathcal{A}(D)||\mathcal{A}(D')) \overset{\Delta}{=} \frac{1}{\alpha -1} \log \big(\mathbb{E}\left[ e^{(\alpha-1)L^{(o)}} \right] \big) \le \rho \alpha 
\end{align*}
Where $\mathcal{D}_\alpha (\mathcal{A}(D)||\mathcal{A}(D'))$ is called $\alpha$-R\'enyi divergence between the distributions of $\mathcal{A}(D)$ and $\mathcal{A}(D')$.
$L^{o}$ is the privacy loss random variable with probability density function $f(x)=\log\frac{\Pr{\mathcal{A}(D)=x}}{\Pr{\mathcal{A}(D')=x}}$.
\end{definition}

The following properties of \zcdp (proven in~\cite{bun2016concentrated}) are useful for our purpose:

\begin{itemize}
    \item \textbf{Gaussian satisfies zCDP.}
    The Gaussian mechanism which answers $f(D)$ with noise $\mathcal{N}(0, \Delta_f ^2 \sigma^2 \mathbf{I})$ satisfies ($\frac{1 }{2\sigma^2}$)-\zcdp.

   \item \textbf{Laplace satisfies zCDP.}
    The Laplace mechanism which answers $f(D)$ with noise $\Lapp{\mathsf{GS}_f x}$ satisfies ($\frac{1 }{2x^2}$)-\zcdp.

    \item \textbf{Linear Composition of zCDP.} 
    If two mechanisms $\mathcal{A}_1$ and $\mathcal{A}_2$ satisfy $\rho_1$-\zcdp and $\rho_2$-\zcdp respectively, their sequential composition  $\mathcal{A} = (\mathcal{A}_1,\mathcal{A}_2)$ satisfies ($\rho_1+\rho_2$)-\zcdp.
    
    \item \textbf{zCDP implies $(\epsilon,\delta))$-DP.}
    If $\AA$ provides $\rho$-\zcdp, then $\AA$ satisfies $(\rho+2\sqrt{\rho\log(1/\delta)},\delta)$-DP for any $\delta>0$.
\end{itemize}

\subsection{Choosing Appropriate Mechanism}

\begin{figure*}[!htpb]
    \centering

    \subfloat[$\epsilon=0.01, \delta=10^{-8}$]{\includegraphics[width=0.45\textwidth]{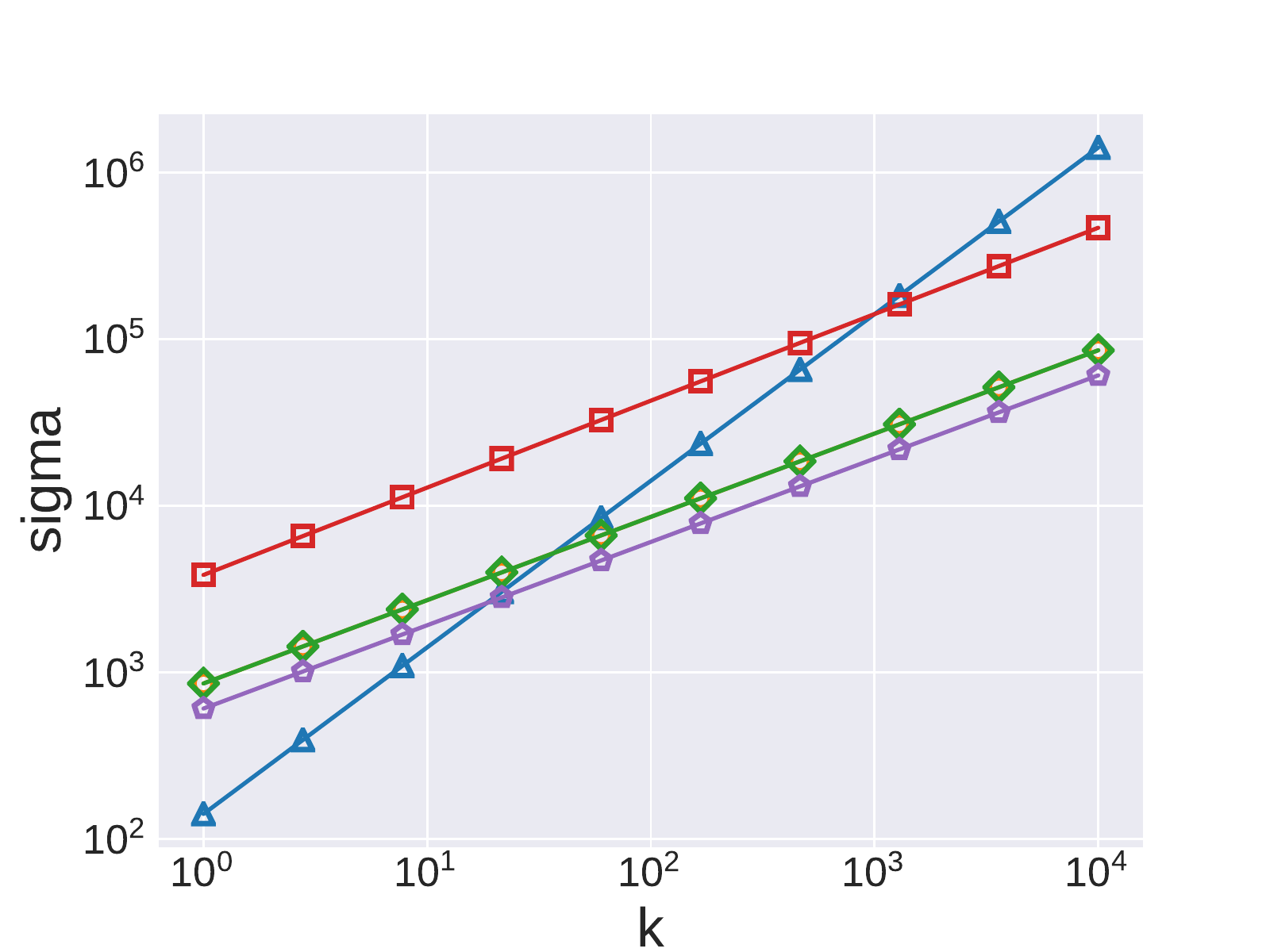}}
    \subfloat[$\epsilon=0.01, \delta=10^{-12}$]{\includegraphics[width=0.45\textwidth]{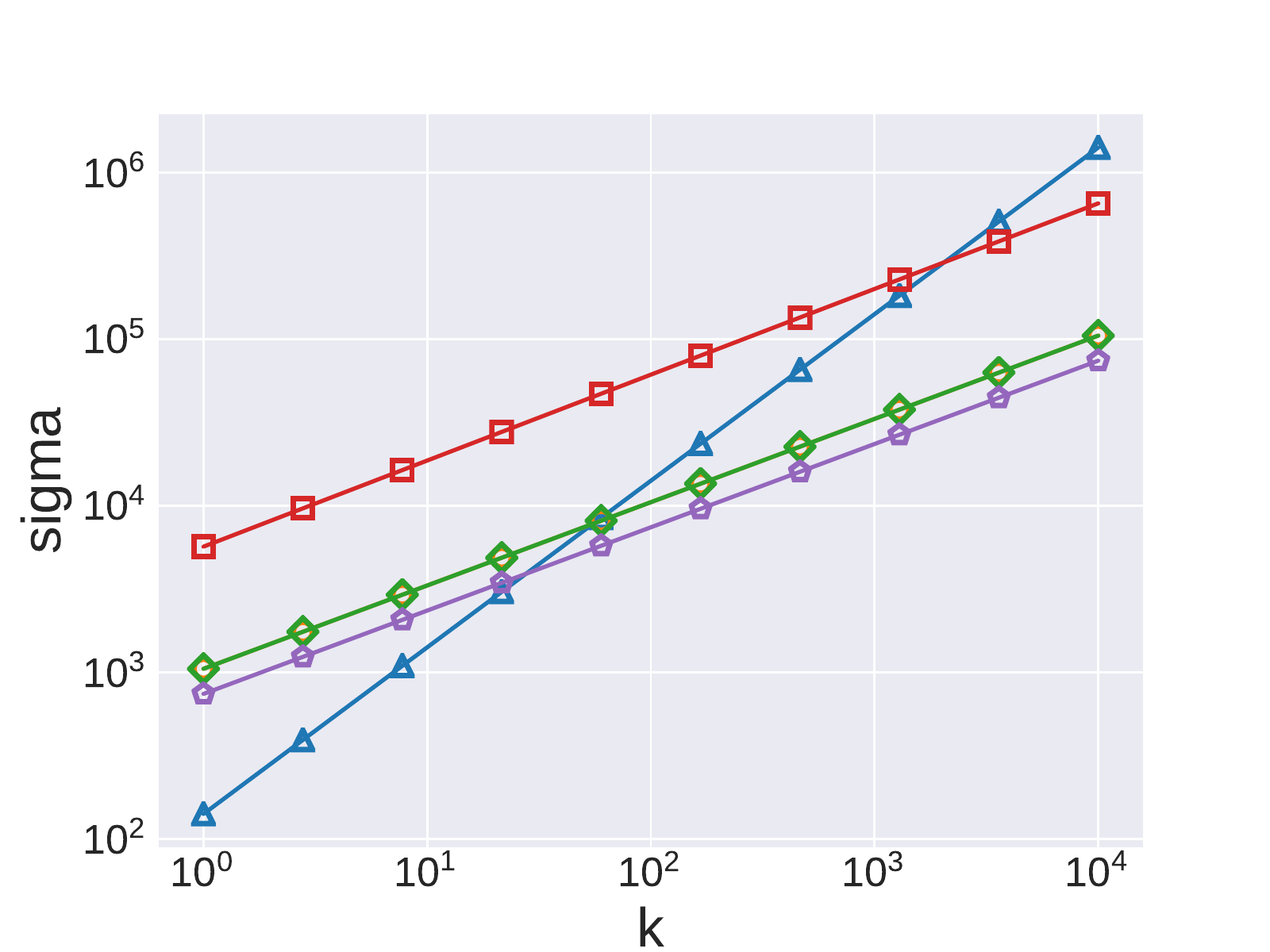}} \\ 
    
    \subfloat[$\epsilon=1.0, \delta=10^{-8}$]{\includegraphics[width=0.45\textwidth]{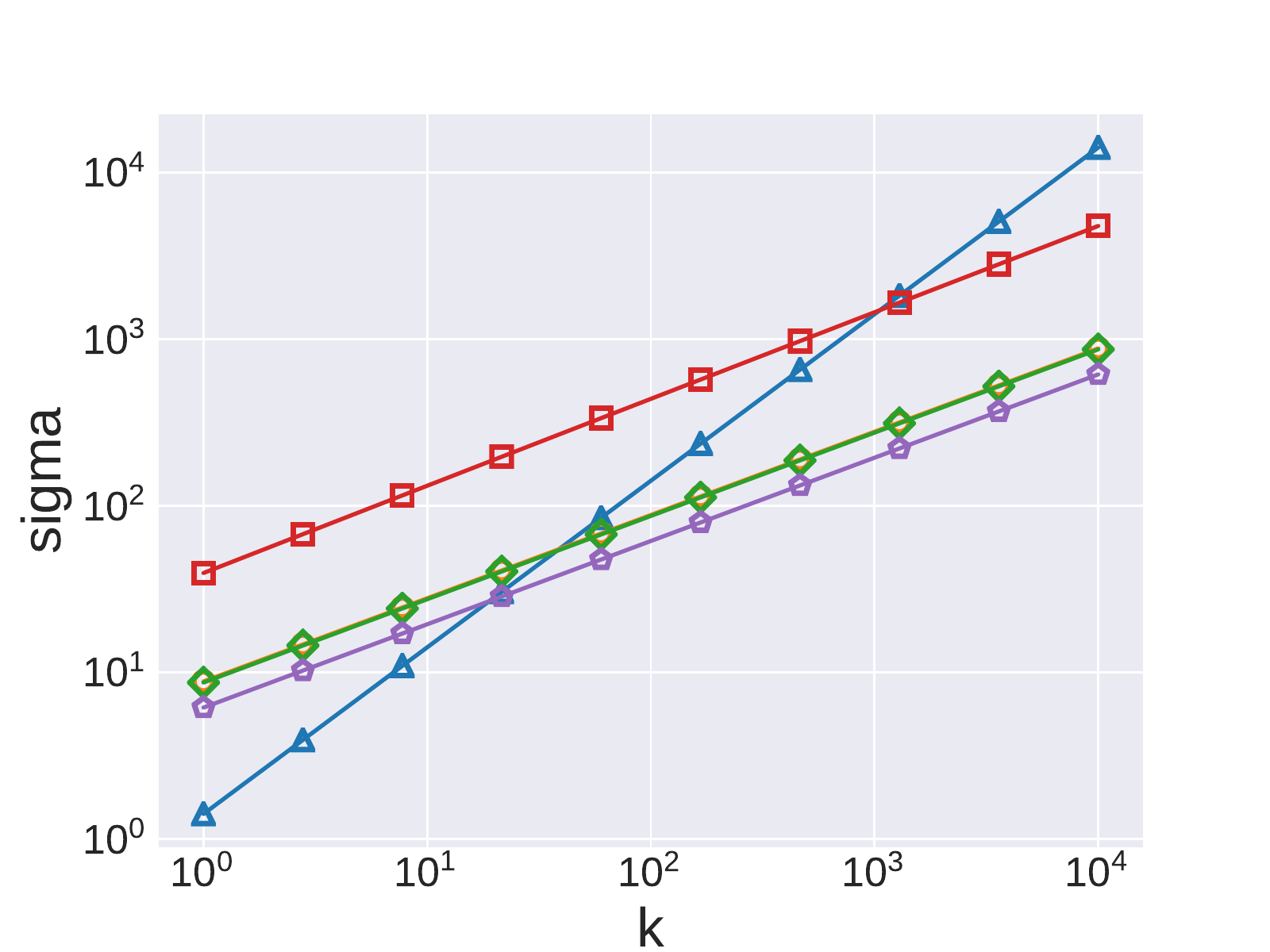}}
    \subfloat[$\epsilon=1.0, \delta=10^{-12}$]{\includegraphics[width=0.45\textwidth]{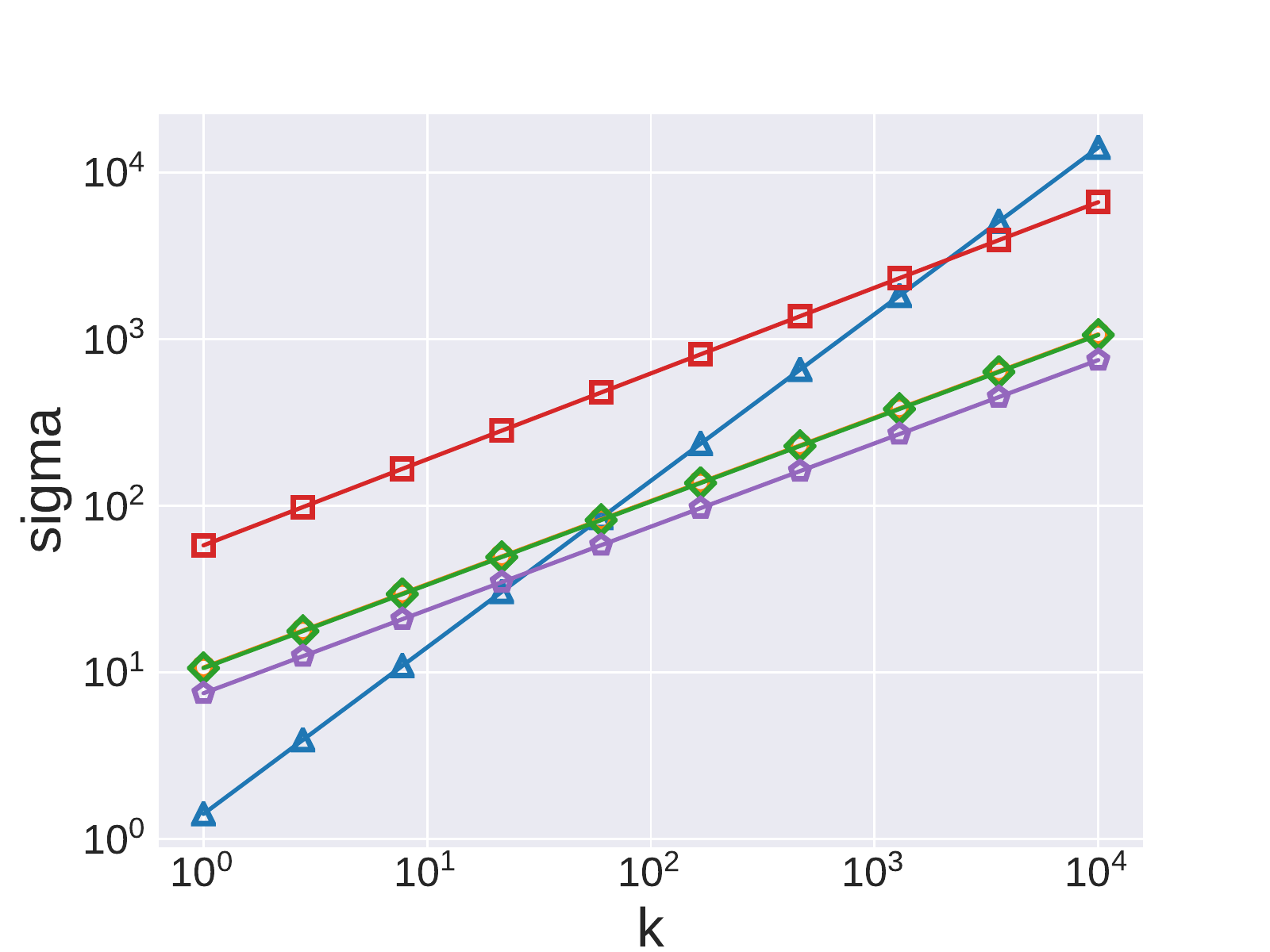}} \\
    
    \subfloat{\includegraphics[width=0.9\textwidth]{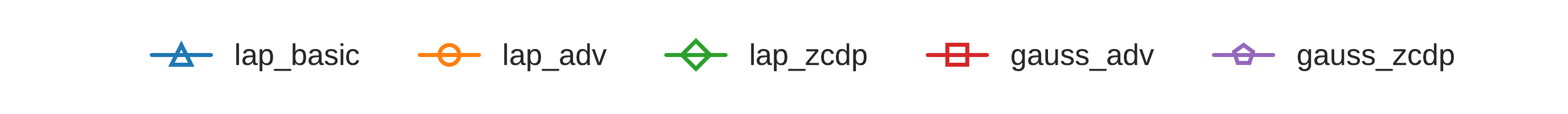}}  \\

    \caption{Comparison of the standard deviation of different composition approaches.
    The standard deviation of ``lap\_adv'' and \revision{``lap\_zcdp''} methods are close to each other (the difference is less than $10$); thus the two lines almost overlap in the log-scale axis.
    The intersection value of $k$ for ``lap\_basic'' and ``gauss\_zcdp'' in four subfigures are (A) 18, (B) 28, (C) 19 and (D) 28.
    }
    \label{fig:composition}
\end{figure*}

Whether one should use the Laplace mechanism or the Gaussian mechanism, and which composition analysis to use depend on several parameters: the privacy parameters $\epsilon$ and $\delta$, and the number of marginals we want to publish, $k$.
Given $\epsilon, \delta$ and $k$, the derivation of the standard deviation for each task can be found in Theorem 7 of~\cite{zhang2021privsyn}.
Figure~\ref{fig:composition} plots the standard deviation of the noises added to each marginal cell under five approaches when $k$ changes.  We can see that when $k$ is small, the best approach is to use Laplace mechanism with basic composition, and when $k$ is large, the best approach is to use zCDP together with the Gaussian mechanism.

\section{Our Experiences before the Competition}
\label{sec:prior}

Before the NIST competition, we have developed state-of-the-art algorithms for several tasks while satisfying $(\epsilon,0)$-DP.  
The experiences and insights we gained in the process played a key role in developing our approach for the NIST competition, and may be helpful for others who need to develop DP algorithms.  Therefore, we describe some of them here in more details.  As this competition uses datasets where the number of attributes are in the tens or close to one hundred, the most relevant work deals with high-dimensional datasets.  

\subsection{The PrivBasis algorithm for Frequent Itemset Mining}

Our first work on developing algorithms for differentially private data analysis is about frequent itemset mining (FIM)~\cite{li2012privbasis}.  In FIM, the dataset consists of a number of transactions, each being a set of items. 
The goal is to find the itemsets that appear in at least $\theta$ fraction of the transactions, and output the frequencies of these itemsets.  As the number of total items can be very large (e.g., from hundreds to thousands), the key challenge here is how to deal with the high dimensionality.   

PrivBasis is based on the following observations.  First, a transaction dataset can be viewed as a relational dataset where each item is a binary attribute.  A record has 1 in an attribute if it contains the corresponding item and 0 otherwise.  Second, if we are able to identify a basis set $\mathcal{V}$ of itemsets such that each frequent itemset is a subset of at least one basis $V \in \mathcal{V}$, then obtaining one marginal for every itemset in $\mathcal{V}$ yields estimated frequencies for all frequent itemsets.  
For example, if $\{a,b,c\}$, $\{b,d\}$, $\{c,e\}$, $\{a,b,g\}$, $\{g,h\}$ are all the maximal frequent itemsets, then $\mathcal{V}=\{V_1=\{a,b,c,d,e\}, V_2=\{a,b,g,h\}\}$ is a basis set.  A marginal for the basis $\mathcal{V}_1=\{a,b,c,d,e\}$ has $2^5$ cells, each corresponding to a subset of $\mathcal{V}_1$.  A transaction $t$ contributes 1 to the cell corresponding to $t\cap \mathcal{V}_1$.  The frequency of, e.g., the itemset $\{a,b,c\}$, can be obtained by summing over $2^2$ cells in the marginal table.   When the size of a basis is large, then the estimation may be noisier, since it needs to sum up more noisy estimations.  On the other hand, when the size of a basis set is large, each basis gets allocated less privacy budget, and the estimation is noisier.  
The main technical challenge in designing the PrivBasis algorithm lies in finding a suitable set $\mathcal{V}$.  
We exploit \revision{the fact that an itemset can be frequent only if all its subsets are}, and repeatedly use the Exponential Mechanism (EM)~\cite{mcsherry2007mechanism} to select frequent items and pairs to construct $\mathcal{V}$.

\subsection{The PriView algorithm for Answering Marginal Queries}

In~\cite{qardaji2014priview}, we tackle the problem of answering arbitrary $k$-way marginal queries in datasets with binary attributes.  \revision{Since PriView is the basis of the DPSyn approach we use in the competition, we describe the PriView approach here.}
Given a $d$-dimension binary dataset $D$, we aim to construct, in a differentially private way, a synopsis of $D$, so that one can relatively accurately compute the marginal table for any set of $k$ attributes, referred to as $k$-way marginals.  We assume that $d$, the number of dimensions, is large so that $\Theta(2^d)$ running time is infeasible.

One baseline approach, which we call the Direct method, is to add independently generated noise to every $k$-way marginal table.  
Since there are \revision{${d\choose k}=O(d^k)$} such marginals, each one is allocated only $1/{d\choose k}$ of the total privacy budget, which is very small when $d$ is large and $k$ is not too small.
\cite{BCD+07} proposed a method of adding noise to the Fourier coefficients that are needed to compute the marginals one wants to compute.  When applying this method to compute all $k$-way marginals, it reduces the error slightly over the Direct method, but still needs to release $O(d^k)$ coefficients.  
Before our work, the problem of generating marginals have been studied in a series of theoretical papers, see, e.g.,~\cite{CKKL12, GHRU11, HR10, HLM12,HRS12,TUV12,DWH+11,LHR+10}; however, these methods do not scale when $2^d$ is large.
In summary, the state-of-the-art before our work is that when $2^d$-complexity is infeasible, then the Fourier method in~\cite{BCD+07} is the best approach, and \revision{it scales poorly due to the need to release $O(d^k)$ coefficients}.  

PriView privately publishes a synopsis of the dataset that takes the form of $m$ size-$\ell$ marginals (i.e., each marginal is specified by $\ell$ different attributes) that are called \emph{views}.  Note that when $\ell$ is large, each marginal covers more attributes; however, estimations computed from the marginals will be noisier.  
\revision{We conducted a heuristic analysis, which showed that choosing $\ell$ to be about $8$ works the best.
The choice of $\ell$ is independent from other parameters such as the dataset size $N$, the number of attributes $d$, and the privacy budget $\epsilon$.
}
The choice of $m$, however, will depends on these parameters, especially $N$ and $\epsilon$.  

In~\cite{qardaji2014priview}, we use the idea of covering design from combinatorial theory~\cite{gordon1995new,coverdesign}, and choose enough size-$8$ views so that every $t$-way marginal is covered in at least one view.  The choice of $t$ depends on $N,d,\epsilon$, and is typically $2$ or $3$.  For example, using covering design one can choose 72 eight-way views (i.e., marginals) to ensure that every $2$-way marginals among 64 attributes are covered in at least one view. 

\begin{example}\label{ex:one}
Given 12 attributes (named from 1 to 12), the following $3$ views cover all pairs ($2$-way marginals). 
$$
\begin{array}{ccc}
    v_1=\{1,2,3,4,5,6,7,8\} & 
    v_2=\{1,2,3,4,9,10,11,12\} & 
    v_3=\{5,6,7,8,9,10,11,12\}
\end{array}
$$
\end{example}
The idea of using covering design in this setting is interesting; however, it turns out to be unnecessary.  Through experiments, we found that one could often achieve similar performance by simply randomly choosing the views when $m$ and $\ell$ are fixed.  

Once these views are selected, we can obtain these marginals for the input dataset by adding sufficient amount of noise to satisfy the desired DP objective.  This is the only step in the PriView algorithm that needs to access the input dataset.  The remaining challenge is how to effectively use these noisy marginals.  Meeting this challenge is the main contributions of~\cite{qardaji2014priview}.  We developed techniques for achieving consistency among the noisy views, and a method to construct arbitrary $k$-way marginals given the views. 

\mypara{Consistency} 
Since different views overlap, one marginal can be derived from different views.  In Example~\ref{ex:one}, one could derive the marginal $\{1,2,3,4\}$ from either $v_1$ or $v_2$.  However, since independent noises are added to $v_1$ and $v_2$, the estimations obtained from them would be different.  In addition, the noisy marginals may contain negative values.  The goal of the consistency step is to ensure that all noisy marginals are non-negative, and whenever two marginals overlap, they are consistent with each other.  This step serves two purposes.  First, it improves the estimation accuracy because averaging over independent perturbations of the same quantity reduces the variance.  Second, consistency enables the next step of constructing arbitrary $k$-way marginals. 

The technique for ensuring two marginals consistent while minimizing variance is known~\cite{HRMS10}.  However, ensuring consistency between one pair of marginals may disturb consistency between other pairs.  Our key insight is that by considering the set of attributes that are included in more than one marginals in the topological sort order (smaller set first), consistency that was already established is never violated later.  For example, in Example~\ref{ex:one}, we first ensure consistency on the empty set, which essentially making all three marginals have the same total count that is the average of their original counts.  After that, one enforces consistency between $v_1$ and $v_2$ on $\{1,2,3,4\}$, consistency between $v_1$ and $v_3$ on $\{5,6,7,8 \}$, and consistency between $v_2$ and $v_3$ on $\{9,10,11,12\}$.  
We also introduce a technique for turning negative values in marginals non-negative while preserving consistency.

\mypara{Generating $k$-way Marginals}
When a $k$-way marginal $q$ is contained in some view, it can be computed from the marginal on that view.  When a $k$-way marginal query $q$ is not fully contained in any of the views, it needs to be estimated.  Any view that overlaps with $q$ provides some information about $q$ that can be viewed as a number of constraints on cells in $q$.  As $q$ is a $k$-way marginal, it has $2^k$ cells, and the constraints from all views form an under-specified system of linear equations on the $2^k$ cell values.  We explored several different methods for computing such a $q$: using linear programming to compute any solution, finding the solution with least $L_2$ norm, and finding the solution with the maximum entropy.  Empirically, we found that the Maximum Entropy approach performs the best.  

\mypara{Discussion}
\revision{
Experimental results in~\cite{qardaji2014priview} show that PriView outperforms the best of previous methods (sometimes by a few orders of magnitude in terms of $L_2$ error), even though it lacks a theoretical utility bound.
}
As of this writing, %
we are unaware of any new method that outperforms PriView for this task.   One main limitation of PriView is that it deals only with binary attributes.

While both PriView and PrivBasis rely on marginals, they differ in a few interesting ways.  PrivBasis can deal with datasets with tens of thousands of attributes/items, because we are interested only in the attributes and their combinations that are frequent.  Thus the key challenge in PrivBasis lies in selecting which marginals to use.  In PriView, we do not assume knowledge regarding which attributes are more interesting to use than others, and thus the choice of marginals is relatively straightforward.  The main challenges lie in how to effectively use the marginals after we obtained them.

\section{DPSyn: Our Response to the Unlinkable Data Challenge}
\label{sec:dpsyn}

The Unlinkable Data Challenge asks for concept papers that propose a mechanism to enable the protection of personally identifiable information while maintaining a dataset's utility for analysis.  We proposed an approach that builds on PriView.  In this section, we describe our original proposed approach.  The text is largely taken from the concept paper\footnote{\url{https://www.herox.com/protected/109/473512/file:DKeSqMQ8rNNfrERUYCFy_2x1y0g}} we submitted to the challenge.  

Our technique deals with categorical datasets; and numerical attributes are first bucketized so that they become categorical values.  Our approach is to randomly select $m$ size-$\ell$ marginals of  (e.g., $m=50, \ell=8$), compute these marginals on the input dataset in a way that satisfies differential privacy, and then synthesize a dataset based on these marginals.  The rationale for our approach is as follows.  Any analysis one may want to conduct on a dataset can be performed using the joint distribution of some subset of attributes.  On most subsets of attributes, a synthesized dataset that simultaneously preserves many (from dozens to hundreds) randomly-chosen marginals would have a distribution close to that of the original dataset.  Under differential privacy, one can answer counting queries with good accuracy, since they have a low sensitivity ($1$ under Unbounded DP).  Each marginal can be viewed as a set of counting queries at the same privacy cost of one counting query.  Thus marginals are probably the most privacy-efficient way of extracting information from the input dataset. 

Given a dataset as input, the first step is to generate many randomly selected noisy marginals on the dataset.  
The algorithm decides the parameters $m$ (number of marginals) and $\ell$, and then computes these marginals of the input dataset, and finally adds Laplacian/Gaussian noise to them so that differential privacy is satisfied.  
\revision{The way PreView chooses $m$ and $\ell$ depends on the fact that all attributes are binary.  Other methods are needed to choose $\ell$ for general categorical attributes.}

In the second step, we use techniques developed in PriView to make all noisy marginals consistent with each other.  The techniques presented in PriView were for binary attributes.  We have already extended those techniques to categorical attributes in~\cite{zhang2018calm}. %
One can use these noisy marginals (called private views in PriView) to reconstruct arbitrary marginals with high accuracy.  This suggests that these noisy marginals capture a lot of information in the input dataset, and can be used for a broad range of data analysis tasks.  

The third step is to generate a synthetic dataset given these private views.  PriView only has techniques to reconstruct arbitrary queried marginal.  Here we propose to develop techniques to synthesize a dataset that approximates these views.  
We plan to investigate a few alternative methods.  One method starts with a randomly generated dataset and gradually changes it to be consistent with the noisy marginals.  Another is to use these noisy marginals to construct probabilistic graphical models of the dataset, and then synthesize data from these probabilistic models.  The key challenge is efficiency.  To be able to preserve information in datasets with dozens or more attributes, we expect to use dozens or more noisy marginals, each including $5-10$ attributes.  We expect the implemented software tool can generate datasets with millions of records in this setting.  

Clearly, the larger $m$ (the more marginals), the more marginal information is preserved, and the better the utility.  However, a larger $m$ also means more noise for each marginal, since the total privacy budget $\epsilon$ is fixed.  Fortunately, a larger dataset means that less privacy budget is needed for each individual marginal.  With a dataset $10$ times larger, one needs only $1/10$ privacy budget to get marginals of the same accuracy.  Furthermore, if we are willing to go beyond strict $\epsilon$-DP, and accepts weaker notions such as ($\epsilon$, $\delta$)-DP, we can use more advanced composition results to our advantage.  Essentially, if one spends $1/10$ of original budget for one marginal, then one is able to publish a lot more than $10$ (close to $10^2$) times the original number of marginals.  Thus the DPSyn approach that relies on marginals to extract information is very promising for large datasets.
\section{Fleshing out DPSyn: Participation in the Synthetic Data Challenge}
\label{sec:setup}

In the Synthetic Data Challenge, we fleshed out the design and implementation of DPSyn.  The main technical development was an algorithm to synthesize a dataset from marginals.  A lot of time during the challenge was spent on designing marginals to achieve high scores according to the metric used in the competition, and the sample dataset. 
Usage of meta information is explicitly allowed because such information is considered public knowledge, and the evaluation is based on hidden datasets that are similar to, but different from the sample dataset.  We believe that this setup illustrates an important aspect of applying DP in practice, when there is often a lot of meta information that can be viewed as public knowledge.  
Later we developed a method to privately, automatically select marginals, and refined the data synthesis algorithm.  These were presented in~\cite{zhang2021privsyn}. 

\subsection{Setup of the Synthetic Data Challenge}
This challenge asks the teams to develop an $(\epsilon, \delta)$-differential privacy algorithm, with $\delta \leq \frac{1}{N^2}$, where $N$ is the number of dataset records, and  the $\epsilon$ values range from 0.1 to 10.

\mypara{Datasets}
The challenge consists of three rounds. 
The first two rounds use the San Francisco Fire Departments Call for Service dataset.  The provided dataset has 305k records and 32 attributes.  
The third challenge uses the Public Use Microdata Sample (PUMS) of the 1940 USA Census Data for the State of Colorado, fetched from the IPUMS USA Website.  The dataset has about 662k records and 98 attributes.

\mypara{Metrics}
Altogether, three metrics were used in the competition.  Round one uses only the first metric; Round two uses both the first and the second; and Round three uses all three metrics.  When multiple metrics are used, they are treated as having equal weights.

\begin{itemize}
    \item \mypara{Density Estimation}
    For this metric, the scoring algorithm randomly samples 300 marginal schemas, each with three randomly chosen attributes.  Then for each marginal schema, compute the normalized marginal tables from the synthetic dataset and from the ground truth dataset, and then use the L1 distance between the two marginal tables as the penalty.  
    Since a normalized marginal table has the property that the sum of all cells in the table is $1$, the penalty is a number $\in [0.0,2.0]$. A difference of 0.0 means a perfect match of density distributions, and a difference of 2.0 means that the distributions for the original and synthetic dataset do not overlap at all. 
    Let $s$ be the average penalty, the resulting score is defined as $S = 10^6 × (1 - s / 2)$, which has a range of $[0,10^6]$.
    
    \item \mypara{Range Query}
    For this metric, the scoring algorithm randomly samples 300 range queries and assesses the accuracy of using the synthetic dataset to answer these queries.  To generate a range query, one first randomly samples a subset of attributes, with each attribute having 33\% chance to be selected.  Then, for each selected attribute, a query condition is randomly generated.  For a categorical attribute, the condition is a randomly picked subset of its possible values (from 1 to maximum number of values).  For a numeric attribute, the condition is a randomly chosen continuous range. 
    A record satisfies the query if and only if in every attribute that is included in the query, the record has a value that is included in the corresponding condition. 

    For $i$-th range query, let $f_{(o,i)}$ be the fraction of records satisfying the query in the original dataset, and $f_{(p,i)}$ be that in the synthetic dataset. 
    When selecting the query, it is guaranteed that in the original dataset there is at least one record that satisfies this query; thus $f_{(o,i)}>0$.  The score is calculate as follows:
    \begin{align*}
        d_i &= \ln{ \frac{\max(f_{(p,i)}, 10^{-6})} {f_{(o,i)}} },  \mbox{ for $i \in [1, 300]$ } \\
        S &= 10^6 \cdot \max \left(0,\;\; 1 -  \left(\frac{1}{\ln{10^{3}}}\right) \cdot \sqrt{\frac{1}{300} \sum_{i \in [1,300]} d_i^2} \right)
    \end{align*}
    Note that both the Density Estimation metric and the Range Query metric are generic, and can be applied to any domain. 

    \item \mypara{Gini Index and Gender Gap} 
    This metric is used only for the IPUMS dataset, and aims to capture specific application needs.  
    For each possible value in the CITY attribute, we calculate, based on the SEX and INCWAGE attributes, Gini index and gender pay gap. 
    We calculate the first score component based on the mean-square deviation between Gini indices obtained for the original and privatized dataset, averaged over the cities present in the original dataset. 
    For the second score component, we rank the cities by the gender pay gap, and calculate the score component based on the mean-square deviation between the resulting city ranks in the original and privatized datasets. 
    The overall score is the average between these two score components, and normalized to the $[0; 1 000 000]$ range.
    
    The three metrics each carries one-third of the weight towards the final score, yet the third metric relies only on three of the 98 attributes.  We thus need to allocate higher privacy budgets to marginals involving these three attributes. 
\end{itemize}

\subsection{Overview of Marginal Engineering}

Given our DPSyn approach, we need to select, for different $\epsilon$ values, the marginal schemas and the privacy budget allocated to each marginal.  We found that which marginals to use has a large impact on the resulting score.  
This is especially true when $\epsilon$ is small. 
We have observed that for larger $\epsilon$ values, the scores are close to the maximal possible score, and there is little room for improvement.  
Thus the focus on marginal engineering is on choosing which marginals to use when the privacy budget is small.   %
We thus need to study the given dataset, and meta-data, and manually decide which configuration to use.

\mypara{Data Exploration}
When given the sample dataset, we examine 1-way marginals on all attributes to get a feeling of the distribution of these attributes.  We also compute a correlation score for every pair of attributes using a metric which we call \margselectfull (\margselect for short).  For any two attributes $a, b$, \margselect calculates the $L_1$ distance between the 2-way marginal $\mathsf{M}_{a,b}$ and 2-way marginal generated from the one-way marginals $\mathsf{M}_a$ and $\mathsf{M}_b$, assuming that $a$ and $b$ are independent, which we use $\mathsf{M}_a \times \mathsf{M}_b$ to denote.  
That is, 
$
\margselect_{a,b}=|\mathsf{M}_{a,b} - \mathsf{M}_a \times \mathsf{M}_b|_1
$.

For pairs of attributes that have low \margselect scores (meaning high correlations), we study their joint distribution to get a sense of the nature of the correlation.  We then look at the meta-data provided in the competition, which provides textual explanation of the attributes and their values.  Almost all correlations can be explained by their semantic meanings. 

\mypara{Parameter Choices}
We need to choose which marginals to use, as well as ways to pre-process attributes.  The decisions often depend on the privacy parameter $\epsilon$, since the competition considers a wide range of $\epsilon$ values. 
When we evaluate the results of different parameter choices, we often compute the accuracy of using synthetic data to estimate all two-way marginals \revision{(we refer the readers to \autoref{sec:synthesizing} for details about the data synthesis)}.
When we discover meta information from the sample dataset, we have to assess whether such information is likely to hold in the evaluation dataset.  
The following two sections discuss our parameter choice for the competition in more detail.

\begin{table}[htbp]
\centering
 \begin{tabular}{c|c|p{9cm}}
\toprule
Attribute Name & \# Values & Definition \\ \midrule
ALS Unit & 2 & Advance Life Support \\ \hline
Final Priority & 2 & Final Call Priority (non-emergency or emergency) \\ \hline
Call Type Group & 5 & Include fire, alarm, potential life threaten and non-life threaten \\ \hline
Original Priority & 8 & Initial Call Priority (non-emergency or emergency) \\ \hline
Priority & 8 & Call Priority (non-emergency or emergency) \\ \hline
City & 9 & City of incident \\ \hline
Unit Type & 10 & Include PRIVATE, MEDIC, ENGINE, CHIEF, TRUCK, SUPPORT, RESCUE CAPTAIN, RESCUE SQUAD, INVESTIGATION, AIRPORT \\ \hline
Fire Prevention District & 11 & Bureau of Fire Prevention District \\
\hline
Battalion & 11 & Emergency Response District \\
\hline
Supervisor District & 12 & Supervisor District \\ \hline
Call Final Disposition & 15 & Disposition of the call \\ \hline
Call Type & 32 & Type of call the incident falls into \\ \hline
Zipcode of Incident & 28 & Zipcode of incident \\ \hline
Neighborhood District & 42 & Neighborhood District \\ \hline
Station Area & 46 & Fire Station First Response Area \\ \hline  
Watch Date & 101 & Watch date when the call is received \\ \hline
Received DtTm & 101 & Date and time received at the 911 Dispatch Center \\ \hline
Entry DtTm & 101 & Date and time the 911 operator submits the entry of the initical call information into the CAD system \\ \hline
Dispatch DtTm & 101 & Date and time the 911 operator dispatches this unit \\ \hline
Response DtTm & 101 & Date and time this unit acknowledges the dispatch \\ \hline
On Scene DtTm & 101 & Date and time the unit records arriving to the location of the incident \\ \hline
Transport DtTm & 101 & If this unit is an ambulance, date and time the unit begins the transport unit arrives to hospital \\ \hline
Hospital DtTm & 101 & If this unit is an ambulance, date and time the unit arrives to the hospital \\ \hline
Location-lng & 101 & Latitude and longitude of address obfuscated either to the midblock, intersection or call box \\ \hline
Number of Alarms & 6 & Number of alarms associated with the incident \\ \hline
Available DtTm & 101 & Date and time this unit is not longer assigned to this call and it is available for another dispatch \\ \hline
Unit sequence & 84 & The order this unit was assigned to this call \\ \hline
Location-lat & 101 & Latitude and longitude of address obfuscated either to the midblock, intersection or call box \\ \hline
Call Date & 101 & Date the call is received at the 911 Dispatch Center \\ \hline
Unit ID & 742 & Unit Identifier \\ \hline
Box & 2089 & Fire box associated with the address of the incident \\ \hline
Address & 17704 & Address of midblock point \\ \bottomrule
 \end{tabular}
 \caption{Description of the Fire Response Dataset.}
 \label{table:fire_dataset}
\end{table}
\subsection{Marginal Engineering for the Fire Response Dataset} 
\label{sec:setup:fire}

The basic goal of designing marginal schemas is to capture as many correlations as possible, while keeping both the number of marginals and the sizes of them (the number of cells) small so that the impact of noise is small.  We developed a few techniques, which could be useful for other domains.  Below we describe these techniques and how we apply them to the Fire Response dataset.  The attributes referred to in the discussions appear in Table~\ref{table:fire_dataset}.  

\mypara{Compress Attributes}  For some attributes, a lot of the values have very low counts.  For such an attribute, we reduce its domain size by merging values with low counts into one single value.  The benefit is that when this attribute is used in a marginal together with other attributes, the number of cells is reduced.  More specifically, after obtaining a noisy one-way marginal for an attribute, we keep marginal cells that have count above a threshold $\theta$.  For the cells that are below $\theta$, we add them up, if their total is still below $\theta$, we assign 0 to all these cells.  If their total is above $\theta$, then we create a new value to represent all values that have low counts.  After synthesizing the dataset, this new value is replaced by the values it represents using the original noisy marginal, assuming a uniform distribution.

The threshold $\theta$ is set using $\theta = \max\left(4.5\sigma, 800\right)$.  
Here $\sigma$ is the standard deviation for Gaussian noises added to the marginal. 
The logic for using several multiples of the standard deviation $\sigma$ is that values below that are likely to be result of adding noises to 0 or very small counts, whereas values above that are much less likely to be just result of noises. 
The number $800$ is loosely related to several parameters, including the number of records and the range of noise standard deviations for two-way marginals.  The latter depends both on the range of privacy parameters $\epsilon,\delta$ and on the number marginals we expect to publish, which depends on the total number of attributes. 
We expect that the algorithm to perform similarly when the value 800 is replaced with a similar one, e.g., 600 or 1000.

\mypara{Recoding Groups of Attributes}  
Some attributes are semantically correlated so that when one attribute takes some other values, the other can take only some values in the domain.  We sometimes recode these attributes into one, so that when they appear in a marginal with other attributes, the domain is smaller.   Below are some examples where we use such group recoding to create new attributes. 

\begin{itemize}
 \item Unit\_Info: Combine ALS Unit and Unit Type.   
 The ALS Unit attribute describes whether the unit includes ALS (Advanced Life Support) resources.  Clearly this is semantically correlated with Unit Type.  

 \item Call\_Type\_Info: Combine Call Type Group and Call Type.
 Call Type Group is a more coarse-grained description of Call Type. 
 
 \item Priority\_Info: Combine Final Priority, Original Priority, and Priority.
 These are all related to the priority of an incident. 
\end{itemize}

\mypara{Date/Time Attributes}  There are 10 date/time attributes in the Fire Response dataset.  In the competition, these time values are binned into 100 buckets.  Since the whole dataset spans from 2000 to 2018, each bucket represent about 2 months.  Since all date/time fields are related to one incidence, we assume that these attributes are synchronous (only a very small number of incidences occur across the boundary and thus have different values in date/time fields).  

However, some of these date/time fields may be missing, e.g., Response DtTm may be missing because there may not be a response, On Scene DtTm may be missing because the unit may not arrive or be cancelled, and so on.  We create a new attribute to identify how many records have missing date/time attributes. 

\begin{itemize}
 \item Time\_Availability: Some date/time attributes may be missing.  We obtain a marginal where the included date/time attributes are encoded in a binary form. For each attribute, there are two possible values: 0 indicates this attribute is missing; and 1 otherwise.  
\end{itemize}

\mypara{Location Coordinate Attributes}  These include Longitude and Latitude.  They have already been binned to 100 values in the given dataset.  How they are handled depends on the available privacy budget. 
 
\begin{itemize}
  \item When $\epsilon$ is low, we just obtain one-way marginals for them, and do not attempt to recover their correlations with other attributes.  

  \item With mid-range $\epsilon$, we recode them into attributes X (for Longitude) and Y (for Latitude), by removing values that have low count. %
  We then obtain the joint distributions of X, Y with other geo-location attributes. 
 
  \item With high-range $\epsilon$, in addition to recoding Longitude and Latitude into X and Y, we also obtain a $100\times100$ 2-way marginal between them, and recode it into a new attribute, for which we obtain joint marginal of this recoded attribute with the Box attribute and the Address attribute. 
\end{itemize}

\mypara{Attributes with Large Domains} They need to be specially handled, because the average counts for each value are low and are easily overwhelmed by noises.  Below we give two examples on how we handle such attributes. 
\begin{itemize}
    \item The Box attribute has 2089 values.  It should be highly correlated with location and other geo-spatial attributes.  When privacy budget is low, such correlation cannot be recovered, and we simply assign random values to this attribute.  When it is high, we use a one-way marginal to encode Box into a new attribute, and obtain a two-way marginal of Box and the coordinate attribute. 

    \item The Address attribute has 17704 values.  It should be highly correlated with other geo-spatial attributes.  When the privacy budget is low, we simply assign random values to this attribute.  When it is high, we use a one-way marginal to encode Address into a new attribute, and obtain a two-way marginal of Box and the coordinate attribute. 
\end{itemize}

\begin{table}[htbp]
\centering
 \begin{tabular}{c|c|p{10cm}}
\toprule
Attribute Name & \#Values & Definition \\ \midrule
SPLIT	&	2	&	Large group quarters that was split up (100\% datasets)	\\ \hline
SEX 	&	2	&	Sex	\\ \hline
CITIZEN	&	6	&	Citizenship status	\\ \hline
EMPSTATD	&	15	&	Employment status [detailed version]	\\ \hline
AGEMONTH	&	15	&	Age in months	\\ \hline
AGE	&	130	&	reports the person's age in years as of the last birthday	\\ \hline
RACED	&	238	&	Race [detailed version]	\\ \hline
DURUNEMP	&	100	&	DURUNEMP is a 2-digit numeric variable that reports how many consecutive weeks had elapsed since each currently-unemployed respondent was last employed (i.e., how many weeks had the person been without a job and looking for one)	\\ \hline
EDUCD	&	44	&	Educational attainment [detailed version]	\\ \hline
METAREAD	&	378	&	Metropolitan area [detailed version]	\\ \hline
CITY	&	1162	& The city in which the person lives		\\ \hline
INCWAGE	&	10000000	&	INCWAGE is a 7-digit numeric code reporting each respondent's total pre-tax wage and salary income - that is, money received as an employee - for the previous year	\\ \bottomrule
 \end{tabular}
 \caption{Description of the attributes in the IPUMS Dataset mentioned in the marginal engineering part.
 The readers can refer to the topcoder website for the full list of attributes.}
 \label{tbl:ipums}
\end{table}

\subsection{Marginal Engineering for the IPUMS Dataset} \label{sec:setup:ipums}
The dataset contains close to 100 attributes.  There are many complex relationships between these attributes.  Table~\ref{tbl:ipums} lists the attributes that are mentioned in the discussions below.

\mypara{Strategy for Attributes}
We handle different attributes using different strategies based on their semantic meanings.  In summary, we have used the following strategies:
\begin{itemize}
    \item For attributes with large domain, we apply several approaches to reduce the domain size.
    \begin{enumerate}
        \item \mypara{Use external metadata} Sometimes, we are able to use public metadata to reduce the domain size.
        For example, the domain size of attribute RACED is $621$ (because its maxval in specs file is $620$, starting from $0$), but according to IPUMS website, the number of legitimate values is $238$.  There are some values that are not used and can be ignored.

        \item \mypara{Compress attributes} 
        \revision{We use the attribute compression technique described in \autoref{sec:setup:fire} to further compress values with estimations below the threshold into a dummy value.}
    
        \item \mypara{Bucketize attributes into bins} For some attributes, we use its bucketized domain to generate marginals with other attributes.  At the same time, we also generate one-way marginals for these attributes using their original value.  When generating the synthetic dataset, the coarse-grained values are used; but after the synthesizing procedure, we use the distribution of the one-way marginal to replace the bucketized value.  
    \end{enumerate}

    \item Generate 1-way marginal to capture the distribution of single attribute, \eg, AGEMONTH, DURUNEMP, CITIZEN, \etc  %
    
    \item Generate 2-way marginals to capture the correlation between two attributes, \eg, AGE and EDUCD, AGE and EMPSTATD, CITY and METAREAD, \etc  No additional one-way marginals are generated for them.
    
    \item Directly fill one attribute based on its mapping relationship with another attribute using the public information.  
    For example, some information is encoded using two attributes: a detailed version (those with ``[detailed version]'' in Table~\ref{tbl:ipums}) and a more general version with fewer values.  Since the general version is determined by the detailed version, and this mapping is public information, we include only the detailed version in the synthesize process, and add the general version after the synthesis step.

    \item Sometimes, two attributes have an approximate one-to-one relationship.  While the fact that such mappings exist can be known based on the semantic meaning, the exact mapping between the values is not in any public meta-data.  For them, we use noisy marginals to figure out the mapping between two attributes; then, fix one attribute and use this mapping to fill another attribute.  %

    \item Because $1/3$ of the score depends on the INCWAGE attribute, we handle it differently.  We use the bucketized value to generate two-way marginals with work-related attributes and three-way marginal with CITY and SEX, and apply different bucketizing scheme for different privacy budget.  
\end{itemize}

\revision{
Because there are over 100 attributes, there are thousands of 2-way marginals.  Obtaining all 2-way marginals would result in too much noises added to each.  We thus divide all attributes into four different groups, where attributes that in different groups have low correlation with each other.  Within each group, we can generate $1$-way, $2$-way or $3$-way marginals among all attributes. 
}
Then, one synthesize four separate datasets and join them, as described below.

\subsection{Synthetic Data Generation}
\label{sec:synthesizing}

Given a set of noisy marginals, the data synthesis step generates a new dataset \ds so that its distribution is consistent with the noisy marginals.  
We initialize a random dataset and iteratively update its records to make it consistent with the marginals.  In each iteration, we go through all marginals, and for each marginal update the dataset based on the marginal.  Using this approach, we can work with a large number of marginals.

In~\cite{zhang2021privsyn} we discuss in detail the different approaches we have tried in the data synthesis, and experimentally compare them.
Here we just give a high-level overview here, and readers are referred to \cite{zhang2021privsyn} for details.

Our approach for updating a dataset based on a marginal resembles the idea of multiplicative  update~\cite{arora2012multiplicative}. As this approach \underline{G}radually \underline{U}pdates \ds based on the \underline{M}arginals; and we call it \gum.  By gradually, we mean that in each iteration we do not update \ds to ensure that it matches the target marginal; instead, the update ensures that the marginal computed from \ds is moving closer to the target. More specifically,  
we use a parameter $\alpha \in (0,1)$, so that for each cell that has a value lower than that in the target marginal, we change the records to increase the cell by $\min\left\{ n^t - n^s, \alpha n^s \right\}$,
where $n^s$ is the number from \ds, and $n^t$ is the number in the target marginal.
That is, each cell will increase by a factor of at most $1+\alpha$.  
In the competition, we used $\alpha=0.2$.  
After we have determined the total increase from all cells that need to be increased, we decrease all cells that need to be decreased by the same proportion so that the total number of records in \ds remains the same.  
For self-containment, we put the details of \gum in \autoref{app:synthesizing}.

\section{Discussions and Related Work}
\label{sec:gap}

There were a series of theoretical results on the hardness of private synthetic data generation, which might appear to rule out the possibility of success in this challenge.  Here we discuss these results, and also the general gap between theory and practice in DP.

\subsection{On Theoretical Impossibility Results}

DPSyn is in the non-interactive setting, in which one publishes a synopsis of the dataset in a way that satisfies DP.  From the synopsis, one can compute answers to queries directly, or generate synthetic data.  This differs from the interactive setting, in which one sits between the users and the database, and answers queries when they are submitted, without knowing what queries will be asked in the future.  

There are a series of negative theoretical results concerning DP in the non-interactive setting ~\cite{DN03,DMT07,DY08,DMNS06}. %
These results have been interpreted to mean that (1) one cannot answer a linear (in the database size $n$) number of queries with small noise while preserving privacy and (2) an interactive approach to private data analysis where the number of queries is limited to be small (sub-linear in the $n$) is more promising.  

The success of the NIST competition suggests that this interpretation does not hold from the empirical perspective.  We think there are two reasons for this.  
First, when the number of dimensions is small compared to $n$, the number of marginals that one needs to query the dataset can be much less than $n$.  
For higher-dimensional dataset, DPSyn and other similar approaches rely on querying a small number of low-dimenional marginals and use them to answer queries.  While one loses theoretical guarantee of correctness, this works well in practice.  
We conjecture that a synthetic dataset that preserves low-degree marginals of an original dataset can be used in place of the original dataset in many settings.  
Second, when one is willing to accept a constant error bound on the normalized marginal table (e.g., each cell has noise with standard deviation $10^{-5}$), the number of marginals that one can answer increases linearly in $n$ under $(\epsilon,0)$-DP, and close to quadratic under $(\epsilon,\delta)$-DP.  The negative theoretical results are for the case where required error bound decreases when $n$ increases. 

\cite{DNR+09} proved that there exist data distributions and classes of counting queries such that there is no efficient algorithm that can synthesize data with provable accuracy bound.  
\cite{UV11} strengthen the result by proving that there exist data distributions such that there is no efficient algorithm that can synthesize data while preserving accurate two-way marginals.  We emphasize that these results do not rule out efficient algorithms that can synthesize data to well approximate 2-way marginals for many data distributions that one is likely to encounter in practice.  
Analogously, that the satisfiability problem is NP-Complete does not preclude the existence of SAT solvers that can solve large SAT instances encountered in practice.  

\cite{Kasiviswanathan_stoc2010} provided perhaps the strongest negative result, as it proves that if all $k$-way marginals are released with error below a certain threshold, one can reconstruct a large fraction of the dataset, violating privacy.  \cite{qardaji2014priview} performed a more in-depth examination of the result and found that the hidden poly-logarithm factors in the results mean that they are irrelevant for practice.  For example, if one considers $6$-way marginals in a dataset with 45 binary attributes and $10^6$ records, the threshold is around $10^{-64}$.  Since publishing $k$-way marginals with error $10^{-5}$ would be sufficiently accurate for almost all tasks, this negative result does not apply.

\subsection{Gap Between Theory and Practice}

\revision{
A paradoxical phenomenon on DP data synthesis in particular, and DP data analysis and publishing in general, is that oftentimes an algorithm that has a theorem proving its utility actually performs poorly in practice.
On the other hand, algorithms that perform well in empirical evaluations tend not to have any meaningful utility theorem. }
We discuss this phenomenon here and challenge the research community to develop techniques and tools that can provide formal utility analysis that is more illuminating about the behavior of algorithm in practice.

\mypara{The Multiplicative Weights Exponential Mechanism (MWEM)}
An illustrating example of this gap is the MWEM~\cite{HLM12}.
This mechanism aims at publishing an approximate of the input dataset $\Data$ so that counting queries in a given set $Q$ can be answered accurately.  
One starts from an uninformative approximation of $\Data$ and iteratively improving this approximation.
In each iteration, one computes answers for all queries in $Q$ using the current approximation, then uses the exponential mechanism to privately select one query $q$ from $Q$ that has the most error, then obtains a new answer to $q(\Data)$ in a way that satisfies the privacy, and finally updates the approximation with this new query/answer pair, using the multiplicative weight update method.  

Given $T$ rounds, a basic version of MWEM keeps all $T+1$ versions of the approximation, and performs a single update with each new query/answer pair.  Finally, for each query it uses the average of the answers obtained from all $T+1$ approximations.
This did not fully utilize the information one obtains from the query/answer pairs.  The more practical method proposed in~\cite{HLM12} uses two improvements.
First, in each round, after obtaining a new query/answer pair, it goes through \textbf{all} known query/answer pairs
$100$ iterations to do multiplicative updates; this causes the approximation to converge to a state that is consistent with all known query/answer pairs.  Second, it uses the last, and almost certainly the most accurate approximation to answer all queries.  

While the two improvements provide much better empirical accuracy, they are done ``at the expense of the theoretical guarantees''~\cite{HLM12}, because a utility theorem is proven for the basic version, which would perform poorly in practice. There is no utility theorem for the improved version, which is the version that actually works and for which empirical evaluation is based.  A large part of the reason is that theoretical analysis can only be applied to algorithms that are simple, e.g., doing only $1$ round of multiplicative update instead of $100$ rounds.  On the other hand, well-performing algorithms are likely difficult to analyze in the traditional framework, because they tend to exploit features that are satisfied by common datasets, but can be violated by pathological examples.

\mypara{Frequent Itemset Mining}
For another example, consider Frequent Itemset Mining (FIM).  In~\cite{BLST10}, a formal definition of utility, $(\delta,\eta)$-useful for FIM, was introduced.  In~\cite{ZNC12}, it is proven that for an $\epsilon$-DP algorithm which is $(\delta,\eta)$-useful for reasonable choice of $\delta,\eta$,
the
$\epsilon$ must be over a certain value.  Given the parameters from commonly used datasets, the bounds on $\epsilon$ have to be
so large that they provide no meaningful privacy.  This, however, does not prevent several algorithms for FIM from performing quite well empirically on datasets commonly used as benchmarks for FIM.

\mypara{Bridging the Gap}
We believe that new theoretical tools are needed to enable more meaningful utility analysis of private algorithms.  
In the current approach, one aims at proving asymptotic worst-case utility guarantees for proposed methods.  This often results in methods that have limited applicability to practical scenarios for a number of reasons.
First, a method with an appealing asymptotic utility guarantee often actually underperforms naive methods except for very large parameters for which applying the method is infeasible or unrealistic in practice.  
Second, the asymptotic analysis ignores constant (and sometimes poly-logarithmic) terms, whereas oftentimes these terms constitute the main difference between competing methods.
Third, practically effective algorithms are sometimes complex, and difficult to analyze, resulting in utility bounds often proven for ``toy'' algorithms.    
Fourth, as the utility guarantee must hold for all datasets (including pathological ones), such guarantees are typically so loose that they are meaningless once we plug in the actual parameters.  Practically effective mechanisms often need to exploit features shared by commonly encountered datasets, and cannot provide meaningful utility guarantees for pathological datasets. 
We hope that some of these challenges can be overcome with the development of new theoretical tools.

\subsection{Additional Related Work}

\cite{BLR08} introduced a theoretical algorithm for generating a synthetic dataset while preserving accuracy on a set of queries.  The algorithm uses the Exponential Mechanism~\cite{mcsherry2007mechanism} to select a dataset among exponentially many possible datasets, using the query accuracy as the quality function.  This algorithm takes exponential time in the number of records in the dataset. 
\cite{BCD+07} proposed a method for the case where the domain size for record is not large, by using linear program to compute a full contingency table that is close to marginals extracted from a dataset.  This method does not scale when the domain is large. 

More practical methods can be classified into three approaches.  
(1) Game Based Methods that formulate the dataset synthesis problem as a zero-sum game~\cite{hardt2012simple,gaboardi2014dual,vietrinew}.
(2) Graphical Model Based Methods use marginals to estimate a graphical model that approximates the distribution of the original dataset in a differentially private way, and include~\cite{zhang2017privbayes,bindschaedler2017plausible,mckenna2019graphical,chen2015differentially}. 
(3) Generative Adversarial Network (GAN)-based approaches that include ~\cite{zhang2018differentially,beaulieu2019privacy,abay2018privacy,frigerio2019differentially,tantipongpipat2019differentially}.
We discussed these approaches in more detail and experimentally compared with the most promising ones among them 
in~\cite{zhang2021privsyn}.

\section{Conclusion and Future Work}
\label{sec:conclusion}

We find the experiences of participating in the competition very rewarding.  The competition forced us to develop and refine our algorithm for synthesizing data from marginals, making it much more efficient and effective than what we originally have.  The competition also demonstrated what is possible under DP.  When each round starts, the scores from all the teams were generally low, but they rapidly increase over time.  The final scores achieved by the top teams were significantly higher than what we thought were possible at the beginning of the competition.  For example, the dataset in Round 3 has 98 attributes and 0.62 million records, making it challenging to preserve information under DP.  This success is in part because for these real-world datasets, there are lots of meta-data information one can exploit to select marginals to achieve better results.  An interesting research direction is to study how to effectively utilize existing public datasets to improve publishing of private datasets. 

\revision{
We believe that by running the competitions, NIST performed a valuable service for the research community and the society at large. The competitions were competently set up and executed, and brought researchers and practitioners together to push the boundary on what is known to be possible regarding publishing synthetic data under differential privacy.  We are looking forward to more of similar efforts in the future. %
We suggest consideration of competitions using randomly synthetic data in future.  When using real datasets, the final results often depend on how well one discovers and utilizes meta-data, more than on how well the data synthesis algorithm works.  A large amount of our time (and likely other teams' time) was spent on marginal engineering, which includes understanding the data schema, finding semantic relationships that we can use to reduce the number of marginals needed and making them smaller.  It would help to have in future a competition that is set up in a way that participants can focus the algorithmic aspects of private data synthesis, instead of how to best use meta-data information.  Perhaps the datasets used in evaluation can be generated by a data generalization process, such as a generative model. 
The data generation process should be public, so that every team automatically has access to the same meta-data information that can be used.  The generation process should have hidden sources of randomness so that the distributions and marginal correlations for generated test datasets are unknown.
}

\section*{Acknowledgments}
We thank the anonymous reviewers for their constructive feedback.  
This work is partially funded by NSFC under grant No. 61731004, U1911401, Alibaba-Zhejiang University Joint Research Institute of Frontier Technologies, the Helmholtz Association within the project ``Trustworthy Federated Data Analytics'' (TFDA) (funding number ZT-I-OO1 4), and United States NSF under grant No.~1931443.

\bibliography{privacy}
\bibliographystyle{abbrvnat}

\revisionstart
\appendix
\section{Data Generation Method from PrivSyn}
\label{app:synthesizing}
To be self-contained, we provide the detailed description of the data generation method from PrivSyn~\cite{zhang2021privsyn}.

Given a set of noisy marginals, the data synthesis step generates a new dataset \ds so that its distribution is consistent with the noisy marginals.  
Existing methods~\cite{zhang2017privbayes,mckenna2019graphical} put these marginals into a graphical model, and use the sampling algorithm to generate the synthetic dataset.  As each record is sampled using the marginals, the synthetic dataset distribution is naturally consistent with the distribution.

The drawback of this approach is that when the graph is dense, existing algorithms do not work.  
To overcome this issue, we use an alternative approach.  
Instead of sampling the dataset using the marginals, we initialize a random dataset and update its records to make it consistent with the marginals.  

\subsection{Strawman Method: Min-Cost Flow (\synflow)}
\begin{figure}[ht]
    \centering

    \includegraphics[width=0.5\textwidth]{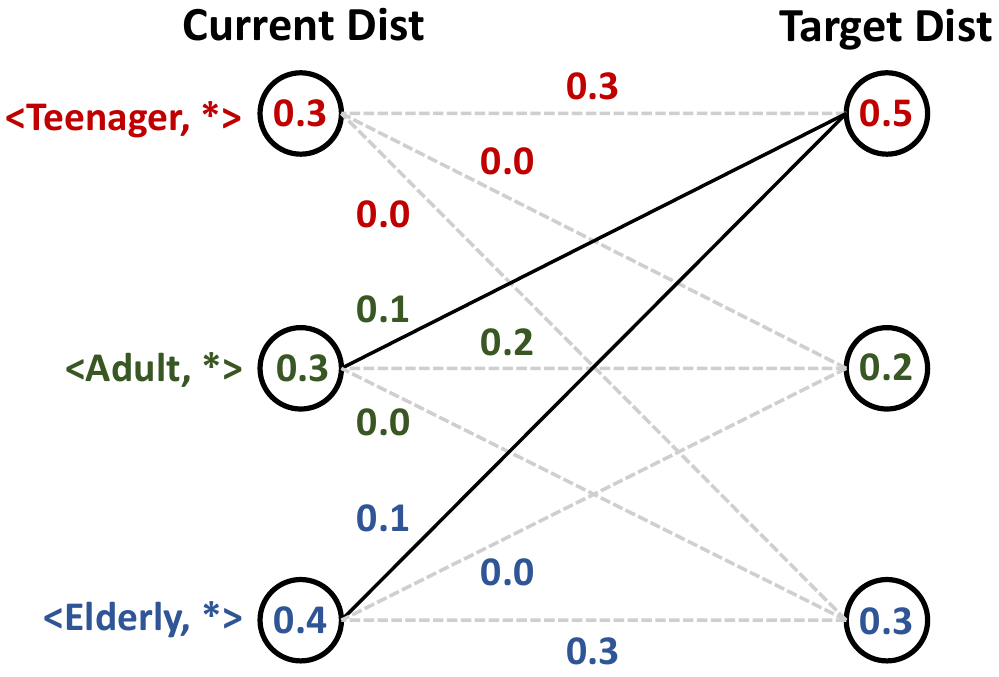}

    \caption{Running example of \synflow.  
    The left nodes represent current distribution from \ds; and the right nodes give the target distribution specified by the noisy marginal.  
    The min-cost flow is to move $0.1$ from adult to teenager, and $0.1$ from elderly to teenager.  
    To change the distribution, we find matching records from \ds and change their corresponding attributes.} 
    \label{fig:mcfp}
\end{figure}

Given the randomly initiated dataset \ds, for each noisy marginal, we update \ds to make it consistent with the marginal.
A marginal specified by a set of attributes is a frequency distribution table for each possible combination of values for the attributes.
The update procedure can be modeled as a graph flow problem.
In particular, given a marginal, a bipartite graph is constructed.  
Its left side represents the current distribution on \ds; and the right side is for the target distribution specified by the marginal.  
Each node corresponds to one cell in the marginal and is associated with a number.  
\autoref{fig:mcfp} demonstrates an example of this flow graph.
Now in order to change \ds to make it consistent with the marginal, we change records in \ds.

The \synflow method enforces a min-cost flow in the graph and updates \ds by changing the values of the records on the flow.
For example, in \autoref{fig:mcfp}, there are two changes to \ds.  
First, one third of the adults needs to be changed to teenagers.  
Note that we change only the related attribute and keep the other attributes the same.  Second, one fourth of the elderly are changed to teenager.  
We iterate over all the noisy marginals and repeat the process multiple times until the amount of changes is small.  
The intuition of using min-cost flow is that, the update operations make the minimal changes to \ds, and by changing the dataset in this minimal way, the consistency already established in \ds (with previous marginals) can be maintained.
The min-cost flow can be solved by the off-the-shelf linear programming solver, e.g.,~\cite{ahuja1988network}.

When all marginals are examined, we randomly shuffle the whole dataset \ds.  
Since the modifying procedure would invalidate the consistency established from previous marginals, \synflow needs to iterate multiple times to ensure that \ds is almost consistent with all marginals.

\begin{figure*}[!htpb]
    \footnotesize
    \centering
	\subfloat[Dataset before updating.]{
		\begin{tabular}[t]{c|c c c}
			\toprule \label{table:dataset}
			& Income  & Gender    & Age  \\ 
			\midrule
			$v_1$     & high          & male          & teenager     \\ 
			$v_2$     & high          & male          & adult     \\ 
			$v_3$     & \brown{high}  & \brown{male}  & \brown{adult} \\ 
			$v_4$     & \brown{high}  & \brown{male}  & \brown{teenager} \\ 
			$v_5$     & \blue{high}   & \blue{female} & \blue{elderly}  \\ 
			\bottomrule
		\end{tabular}
	}
	\hspace{5mm}
	\subfloat[Marginal table for $\{$Income, Gender$\}$,
	where red and blue stands for over-counted and under-counted cells, respectively.
	]{
		\begin{tabular}[t]{c|c c}
			\toprule \label{table:m1}
			$v$		&  $\mathsf{S_{\{\mbox{I,G}\}}}(v)$ & $\mathsf{T_{\{\mbox{I,G}\}}}(v)$ \\ 
			\midrule
			$\langle$low, male,$*\rangle$    & 0.0   & 0.0    \\ 
			$\langle$low, female,$*\rangle$  & 0.0   & 0.0   \\
			$\langle$high, male,$*\rangle$   & \red{0.8}   & \red{0.2}    \\ 
			$\langle$high, female,$*\rangle$ & \blue{0.2}  & \blue{0.8}   \\ 
			\bottomrule
		\end{tabular}
	} 
	\hspace{5mm}
	\subfloat[Dataset after updating.]{
		\begin{tabular}[t]{c|c c c}
			\toprule \label{table:dataset}
			& Income  & Gender    & Age   \\ 
			\midrule
			$v_1$   & high           & male      & teenager     \\ 
			$v_2$   & high           & male      & adult     \\ 
			$v_3$   & \blue{high}   & \blue{female}  & \blue{elderly} \\ 
			$v_4$   & \blue{high}   & \blue{female}  & \brown{teenager} \\ 
			$v_5$   & \blue{high}   & \blue{female}  & \blue{elderly}  \\ 
			\bottomrule
		\end{tabular}
	}
    \caption{Example of the synthesized dataset before and after updating procedure.  
    In (a), blue stands for the records to be added, and brown stands for the records to be changed.
    In (c), $v_4$ only changes income and gender attributes, while $v_3$ changes the whole record which is duplicated from $v_5$.
    Notice that in this example, we have $\alpha=2.0, \beta=0.5$ and the marginal distribution in (c) do not completely match $\mathsf{T_{\{\mbox{I,G}\}}}(v)$ of $[0.0, 0.0, 0.2, 0.8]$; instead, it becomes $[0.0, 0.0, 0.4, 0.6]$.
    }
    \label{fig:update_example}
\end{figure*}

\subsection{Gradually Update Method (\gum)}
\label{subsec:gum}
Empirically, we find that the convergence performance of \mcf is not good.  
We believe that this is because \mcf always changes \ds to make it completely consistent with the current marginal in each step.
Doing this reduces the error of the target marginal close to zero, but increases the errors for other marginals to a large value.

To handle this issue, we borrow the idea of multiplicative  update~\cite{arora2012multiplicative} and propose a new approach that \underline{G}radually \underline{U}pdate \ds based on the \underline{M}arginals; and we call it \gum.  
\gum also adopts the flow graph introduced by \mcf, but differs from \mcf in two ways:  First, \gum does not make \ds fully consistent with the given marginal in each step.  
Instead, it changes \ds in a multiplicative way, so that if the original frequency in a cell is large, then the change to it will be more.  
In particular, we set a parameter $\alpha$, so that for cells that have values are lower than expected (according to the target marginal), we add at most $\alpha$ times of records, i.e., $\min\left\{ n^t - n^s, \alpha n^s \right\}$~\footnote{Notice that $\alpha$ could be greater than $1$ since $n^s < n^t$. 
In the experiments, we always set $\alpha$ to be less than $1$ to achieve better convergence performance.}, where $n^t$ is the number in the marginal and $n^s$ is the number from \ds.
On the other hand, for cells with values higher than expected, we will reduce $\min\left\{ n^s - n^t, \beta n^s \right\}$ records that satisfy it. 
As the total number of record is fixed, given $\alpha$, $\beta$ can be calculated.

\autoref{fig:update_example} gives a running example.  
Before updating, we have 4 out of 5 records have the combination $\langle high, male \rangle$, and 1 record has $\langle high, female \rangle$.  
To get closer to the target marginal of 0.2 and 0.8 for these two cells,
we want to change 2 of the $\langle high, male \rangle$ records to be $\langle high, female \rangle$.
In this example, we have $\alpha=2.0, \beta=0.5$~\footnote{\revision{}{We have $\alpha=\frac{n^t-n^s}{n^s}$ for under-counted cells and $\beta=\frac{n^s-n^t}{n^s}$ for over-counted cells.
The number of records for under-counted cell $\langle$high, female,$*\rangle$ increase from $1$ to $3$; thus $\alpha=\frac{3-1}{1}=2$.
The number of records for over-counted cell $\langle$high, male,$*\rangle$ decrease from $4$ to $2$; thus $\beta=\frac{4-2}{4}=0.5$.}}
and do not completely match the target marginal of 0.2 and 0.8.
To this end, one approach is to simply change the Gender attribute value from male to female in these two records as in \mcf.  
We call this a {\bf Replace} operation.  Replacing will affect the joint distribution of other marginals, such as $\{Gender, Age\}$.
An alternative is to discard an existing $\langle high, male \rangle$ record, and {\bf Duplicate} an existing $\langle high, female \rangle$ record (such as $v_5$ in the example).  
Duplicating an existing record help preserve joint distributions between the changed attributes and attributes not in the marginal.  
However, Duplication will not introduce new records that can better reflect the overall joint distribution.  
In particular, if there is no record that currently has the combination $\langle high, female, elderly \rangle$, duplication cannot be used.  

Therefore, we need to use a combination of Replacement and Duplication (which is the case in \autoref{fig:update_example}).
Furthermore, once the synthesized dataset is getting close to the distribution, we would prefer Duplication to Replacement, since at that time there should be enough records to reflect the distribution and Replacement disrupts the joint distribution between attributes in a marginal and those not in it.  
We empirically compare different record updating strategies and validate that introducing the Duplication operation can effectively improve the convergence performance.

\subsection{Improving the Convergence}
\label{subsec:improve_convergence}
Given the general data synthesize method, we have several optimizations to improve its utility and performance.
First, to bootstrap the synthesizing procedure, we require each attribute of \ds follows the 1-way noisy marginals when we initialize a random dataset \ds.  

\mypara{Gradually Decreasing $\alpha$}
The update rate $\alpha$ should be smaller with the iterations to make the result converge.
From the machine learning perspective, gradually decreasing $\alpha$ can effectively improve the convergence performance.  
There are some common practices~\cite{decay} of setting $\alpha$.

\begin{itemize}
    \item Step decay: $\alpha = \alpha_0 \cdot k^{\lfloor \frac{t}{s} \rfloor}$, where $\alpha_0$ is the initial value, $t$ is the iteration number, $k$ is the decay rate, and $s$ is the step size (decrease $\alpha$ every $s$ iterations).
    The main idea is to reduce $\alpha$ by some factor every few iterations.  
    
    \item Exponential decay: $\alpha = \alpha_0 \cdot e^{-kt}$, where $k$ is a hyperparameter.
    This exponentially decrease $\alpha$ in each iteration. 
    
    \item Linear decay: $\alpha = \frac{\alpha_0}{1+kt}$.
    
    \item Square root decay: $\alpha = \frac{\alpha_0}{\sqrt{1+kt}}$.
    
\end{itemize}

We empirically evaluate the performance of different decay algorithms and find that step decay is preferable in all settings.
The step decay algorithm is also widely used to update the step size in the training of deep neural networks~\cite{krizhevsky2012imagenet}.

\mypara{Attribute Appending}
The selected marginals $\mathcal{X}$ can be represented by a graph $\mathcal{G}$.
We notice that some nodes have degree $1$, which means the corresponding attributes are included in exactly one marginal.
For these attributes, it is not necessary to involve them in the updating procedure.  Instead, we could append them to the synthetic dataset \ds after other attributes are synthesized.
In particular, we identify nodes from $\mathcal{G}$ with degree $1$.  We then remove marginals associated with these nodes from $\mathcal{X}$.  The rest of the noisy marginals are feed into \gum to generate the synthetic data but with some attributes missing.  For each of these missed attributes, we sample a smaller dataset \ds' with only one attribute, and we concatenate \ds' to \ds using the marginal associated with this attribute if there is such a marginal; otherwise, we can just shuffle \ds' and concatenate it to \ds.
Note that this is a one time operation after \gum is done. No synthesizing operation is needed after this step.

\mypara{Separate and Join}
We observe that, when the privacy budget is low, the number of selected marginals is relatively small, and the dependency graph is in the form of several disjoint subgraphs.
In this case, we can apply \gum to each subgraph and then join the corresponding attributes.
The benefit of Separate and Join technique is that, the convergence performance of marginals in one subgraph would not be affected by marginals in other subgraph, which would improve the overall convergence performance.

\mypara{Filter and Combine Low-count Values}
If some attributes have many possible values while most of them have low counts or do not appear in the dataset.
Directly using these attributes to obtain pairwise marginals may introduce too much noise.
To address this issue, we propose to filter and combine the low-count values.
The idea is to spend a portion of privacy budget to obtain the noisy one-way marginals.
After that, we keep the values that have count above a threshold $\theta$.  
For the values that are below $\theta$, we add them up, if the total is below $\theta$, we assign 0 to all these values.  
If their total is above $\theta$, then we create a new value to represent all values that have low counts. 
After synthesizing the dataset, this new value is replaced by the values it represents using the noisy one-way marginal. 
The threshold is set as $\theta = 3 \sigma$, where $\sigma$ is the standard deviation for Gaussian noises added to the one-way marginals.
\revisionend

\end{document}